\documentclass{epl}
\usepackage{amstext,amsmath,amssymb}
\usepackage{mymacros}
\title{
Solving quantum master equations in phase space\\
by con\-tinued-fraction methods
}
\shorttitle{solving quantum master equations in phase space\ldots}
\author{
J. L. Garc\'{\i}a-Palacios
}
\institute{
Dep.\ de F\'{\i}sica de la Materia Condensada e
Instituto de Ciencia de Materiales de Arag\'on,
C.S.I.C.--Universidad de Zaragoza,
E-50009 Zaragoza, Spain
}
\pacs{03.65.Yz}{Decoherence; open systems; quantum statistical
methods}
\pacs{05.40.-a}{Fluctuation phenomena, random processes, noise, and
Brownian motion}
\pacs{05.60.-k}{Transport processes}

\begin{document}

\maketitle

\begin{abstract}
Inspired on the continued-fraction technique to solve the classical
Fokker--Planck equation, we develop continued-fraction methods to
solve quantum master equations in phase space (Wigner representation
of the density matrix).
The approach allows to study several classes of nonlinear quantum
systems subjected to environmental effects (fluctuations and
dissipation), with the only limitations that the starting master
equations may have.
We illustrate the method with the canonical problem of quantum
Brownian motion in periodic potentials.
\end{abstract}

\section{Introduction}
The phase-space formulation of quantum mechanics \cite{hiletal84} has
received a renewed attention in the last decades because it allows to
employ notions and tools of classical physics in the quantum realm.
The central object in this formulation is the Wigner function
$\Wf(\x,\p)$, which is a phase-space representation of the density
matrix $\rho(\x,\x')$.
In a closed system, the equation governing its time evolution is
equivalent to the Schr\"odinger equation, while in the classical limit
it reduces to the Liouville equation for the phase-space distribution.

In an open system, i.e., in contact with the surrounding medium, the
system experiences generically dissipation, fluctuations, and
decoherence \cite{weiss}.
Thus, the study of quantum open systems is of interest in many
areas of physics and chemistry.
Typically one is interested in a system consisting of a few relevant
degrees of freedom coupled to its environment, or bath, which has a
very large number of degrees of freedom.

Under certain conditions, the dynamics of an open system can be
formulated in terms of a quantum master equation for the (reduced)
density matrix \cite{kargra97,ank2003qme}.
In the Wigner representation the master equation goes in the classical
limit over the Klein--Kramers equation 
(Fokker--Planck equation in phase space \cite{risken}).
Recall that phase-space dynamics also includes problems reducible to
``mechanical'' analogues: Josephson junctions, certain electrical
circuits, chemical reactions, etc.
The most powerful technique to solve the Klein--Kramers equation is
the continued-fraction method.
The distribution is expanded in a basis and the equations
for the expansion coefficients derived (moment equations).
These equations, the so-called Brinkman hierarchy, have a tridiagonal
structure that can be exploited to find the solution with continued
fractions \cite{risken}.
The method has also been successfully extended to rotational Brownian
motion problems of classical spins and dipoles (see, for
instance, \cite{kalcof97} and references therein).

To deal with quantum dissipative systems, however, is a more complex
task.
Quantum master equations, except in some simple cases, are difficult
to solve.
Exact expressions for the time evolution of the density matrix
obtained by path-integral methods are available \cite{grasching88}.
However, they are in general difficult to evaluate, even numerically,
because the propagating function is highly oscillatory.
Finally, quantum Monte Carlo simulations can in principle be used, but
they are computationally complex and suffer from the (dynamical) sign
problem.

This situation constitutes a strong motivation for the development of
alternative methods for quantum dissipative systems, even if they are
valid only for specific classes of problems.
Restricting our attention to the quantum master equation approach,
Shibata and co-workers \cite{shi80} developed continued-fraction
methods for a quantum spin in contact with a thermal bath.
Vogel and Risken \cite{vogris88}  employed similar techniques in
quantum nonlinear optics.
However, to the best of our knowledge, this method has not been tried
for the quantum master equations in genuine phase-space problems.

In this article we derive the hierarchy of moment equations
corresponding to a generic master equation for the Wigner function
of the Caldeira--Leggett type \cite{calleg83pa}.
Then we show that, in various cases of interest, the
continued-fraction method for the classical problem can be adapted to
solve this quantum hierarchy, without a sizable increase in
sophistication or computational complexity.
This results in a suitable nonperturbative technique to study
nonlinear quantum systems subjected to dissipation and thermal
fluctuations.
Finally, we illustrate the method with the problem of quantum Brownian
motion in periodic potentials.

\section{Quantum master equations in phase space}
We shall consider the following generic master equation for the Wigner
function of a particle of mass $\sM$ in a potential $\V(\x)$
\cite{calleg83pa,anahal95}:
%
\begin{equation}
\label{WKK:Dxp}
\pt
\Wf
=
\bigg[
-
\frac{\p}{\sM}\,
\px
+
\V'\,
\pp
+
\pp
\big(
\gamma\,\p
+
\Dpp
\pp
\big)
+
\Dxp\pxp
+
\sum_{\iq=1}^{\infty}
\frac{(\iu\hbar/2)^{2\iq}}{(2\iq+1)!}
\V^{(2\iq+1)}
\,
\pp^{(2\iq+1)}
\bigg]
\Wf
\end{equation}
The first two terms (Poisson bracket) generate the classical
reversible evolution (Liouville equation).
The next accounts for irreversible effects due to the coupling to the
environment ($\gamma$ and $\Dpp$ are the dissipation and diffusion
coefficients).
The mixed diffusion term $\Dxp\pxp\Wf$ is heuristically related with the
colour of the quantum noise.
Finally, the infinite series of derivatives $\pp^{(2\iq+1)}\Wf$ gives
the quantum contribution to the unitary evolution of the closed
system.
Conditions under which this series can be truncated are sometimes
discussed, implicitly assuming that solving an infinite-order partial
differential equation would be a hopeless task.
We shall however keep these terms, which can be specially important in
nonlinear systems \cite{zurpaz94}.
To conclude, in the phase-space formulation, the quantum-mechanical
average of any operator $\hat{A}$ is simply obtained from the
corresponding classical variable $A(\x,\p)$ as
%
\begin{equation}
\label{average}
\big\langle A\big\rangle
\equiv
\mathrm{Tr}(\hat{\rho}\,\hat{A})
=
\int\!\drm{\x}\drm{\p}\,
\Wf(\x,\p)\,
A(\x,\p)
\;,
\end{equation}
i.e., multiplying $A(\x,\p)$ by the ``distribution'' $\Wf(\x,\p)$ and
integrating over the phase space.

Conditions of validity for the description in terms of quantum master
equations are discussed in detail in \cite{kargra97,ank2003qme}.
We only mention that equations of the type of eq.~(\ref{WKK:Dxp}) are
usually derived under assumptions like semiclassical
(high-temperature) bath, weak system-bath coupling, etc.
Nevertheless, the quantum master equation recently derived for strong
coupling, and valid for all $\hbar\gamma/\kT$ \cite{ank2003epl},
involves terms {\em structurally\/} similar to eq.~(\ref{WKK:Dxp})
(with $\x$-dependent coefficients).
In addition, the phase-space representation of the celebrated Lindblad
master equation \cite{isasansch96} is obtained by simply adding to
eq.~(\ref{WKK:Dxp}) terms of the form $\px(x\,\Wf)$ and $\px^{2}\Wf$.
All these types of terms can be readily incorporated in the treatment
below.

We shall use scaled variables based on a reference length $\xo$ and a
frequency $\wo$ (e.g., energy scaled by $\Eo=\sM\,\wo^{2}\xo^{2}$,
action by $\So=\sM\,\wo\xo^{2}$, etc.).
Besides, $\Dpp$ ($=\gamma\sM\kT$ in a high-$T$ bath) is handled as an
effective temperature $\kT_{\eff}=\Dpp/\gamma\sM$ and $\p$ is rescaled
by $(\sM\kT_{\eff})^{1/2}$.
Then $\V(\x)$ enters divided by $\kT_{\eff}$ and $\gamma$ as
$\gammaT=\gamma(\sM\xo^{2}/\kT_{\eff})^{1/2}$.
Finally $\hbar$ is introduced as $\hbar/\So=2\pi/\kondobar$, being
$\kondobar$ related to the ordinary Kondo parameter by
$\kondo=(\gamma/\wo)\kondobar$.
$\ldB=\pi\gammaT/\kondo$ is the thermal de Broglie wave length (in
units of $\xo$).

\section{Derivation of the quantum  hierarchy of moment equations} 
In the expansion into complete sets approach to solve kinetic
equations in non-equilibrium statistical mechanics
\cite[p.~175]{balescu2}, the distribution is expanded in an
orthonormal basis $\{\psi_{\ip}\}$ and the equations of motion for the
coefficients $\ec_{\ip}$ derived.
In our case we expand $\Wf$ into Hermite functions $\psi_{\ip}(\p)$
%
\begin{equation}
\label{W:expansion}
\Wf(\x,\p)
=
\Wo
\sum_{\ip}
\ec_{\ip}(\x)
\psi_{\ip}(\p)
\;,
\qquad
\Wo
=
\frac{\e^{-\eta\,\p^{2}/2}}{(2\pi)^{1/4}}\,
\e^{-\bVo(\x)}
\;,
\qquad
\psi_{\ip}
=
\frac
{\e^{-\p^{2}/4}H_{\ip}(\p/\sqrt{2})}
{\sqrt{(2\pi)^{1/2}2^{\ip}\ip!}}
\;.
\end{equation}
The Boltzmann-like factor $\Wo$ is extracted for mathematical
convenience and involves the auxiliary parameter $0\leq\eta\leq1/2$
and the effective potential $\bVo(\x)$ [usually a scaled version of
$\V(x)$].
The Hermite basis has a number of advantages \cite{risken}; not the
least is the handling of the derivatives $\pp^{(\iq)}$ by means of the
associated ``creation'' and ``annihilation'' operators
$\bp=-\pp+\half\p$ and $\bd=\pp+\half p$, which obey the ordinary
boson commutation rule $\big[\bd,\bp\big]=1$

From the orthonormality of the $\{\psi_{\ip}\}$ we have
$\ec_{\ip}(\x,t)=\int\!\drm \p\,\psi_{\ip}\,(\Wf/\Wo)$.
Differentiating with respect to $\tT$ and using the master equation
$\ptT\Wf=\Ls\,\Wf$, one gets the dynamical equations for the
``coefficients'' in the form
$\ptT\ec_{\ip}=\sum_{\ipp}\Q_{\ip\ipp}\ec_{\ipp}$.
To get the (operator) matrix elements
$\Q_{\ip\ipp}
=
\int\!\drm \p\,\psi_{\ip}\big(\Wo^{-1}\Ls\,\Wo\big)\psi_{\ipp}$
one takes advantage of results for normal ordering of
$(\bp\pm\bd)^{\iq}$ \cite{pat2000}.
This leads to the following {\em quantum (Brinkman) hierarchy}
%
\begin{eqnarray}
\label{QBH}
-\ptT
\ec_{\ip}
&=&
\sum_{\iq=0}^{[(\ip-1)/2]}
\,
\big[
\G_{\ip}^{\iq,-}
\,
\bV^{(2\iq+1)}
\big]
\,
\ec_{\ip-(2\iq+1)}
+
\sqrt{(\ip-1)\ip}\;
\gammamm
\,
\ec_{\ip-2}
\nonumber\\
& &
{}+
\sqrt{\ip}\;
\Dm
\ec_{\ip-1}
+
\gammao^{\ip}
\,
\ec_{\ip}
+
\sqrt{\ip+1}\;
\Dp
\ec_{\ip+1}
\nonumber\\
& &
{}+
\sqrt{(\ip+1)(\ip+2)}\;
\gammapp
\,
\ec_{\ip+2}
+
\sum_{\iq=0}^{\infty}
\big[
\G_{\ip+(2\iq+1)}^{\iq,+}
\,
\bV^{(2\iq+1)}
\big]
\,
\ec_{\ip+(2\iq+1)}
\;.
\end{eqnarray}
The auxiliary parameters introduced are combinations of $\gammaT$,
$\eta$, $\etapm=\eta\mp1/2$, and $\etapro=\etam\etap$:
%
\begin{equation}
\label{gammas}
\gamma_{\pm}
=
\gammaT\,
\etapm(1-\etapm)
\;,
\qquad
\gammao^{\ip}
=
\gammaT\,
\left[
(\etap-\etapro)
+
2\ip
(\eta-\etapro)
\right]
\;.
\end{equation}
The operators on the $\x$ dependence read [the $\G$'s act only on
$\V(\x)$ not on the $\ec_{\ipp}(\x)$'s]
%
\begin{eqnarray}
\label{DmDp}
\Dpm
&=&
\dplmi
\big(\px-\bVo'\big)
\;,
\qquad
\dplmi
=
1+\etapm\bDxp
\;,
\qquad
\dlpxx
=
\etapro\ldB^{2}\px^{2}
\;,
\\
\label{Gamma:qcoeff}
\G_{\ip}^{\iq,\pm}
&=&
\etapm^{2\iq+1}
\qcoef^{(\iq)}_{\ip}
\,
\e^{-\dlpxx/2}
\kum\big(-\ipp,2\iq+2\,;\dlpxx\big)
\;,
\quad
\qcoef^{(\iq)}_{\ip}
=
\frac{(-1)^{\iq}\ldB^{2\iq}}{(2\iq+1)!}
\sqrt{\dfrac{\ip!}{\ipp!}}
\;,
\end{eqnarray}
being $\kum(a,c\,;z)$ the confluent hypergeometric (Kummer) function and
$\ipp=\ip-(2\iq+1)$.

In the classical case only the $\iq=0$ terms survive in the above
sums; if we set the usual $\eta=1/2$, eq.~(\ref{QBH}) reduces to the
$3$-term Brinkman hierarchy associated to the classical Klein--Kramers
equation \cite{risken}.
However, except for polynomial potentials $\V^{(\iq)}\equiv0$,
$\forall\iq\ge S$ (harmonic oscillator, Duffing oscillator, etc.), the
quantum terms lead to an infinite range of coupling in the index
$\ip$.
This prevents the use of the custom continued-fraction method (based
on the $\ip$-recurrence).
In the following we shall exploit the expansion in the position basis
to show how this problem can be circumvented.

\section{Expansion into the position basis}
The coefficients $\ec_{\ip}$ are still functions of $\x$.
By expanding them in an orthonormal basis $\{u_{\ix}(\x)\}$ (to be
specified for a given problem), we get a two index recurrence
involving matrix elements of the form
$A_{\ix\ixp}=\int\!\drm{\x}\,u_{\ix}^{\ast}\,A(\x,\px)\,u_{\ixp}$
%
\begin{equation}
\label{dcdt:generic}
\ec_{\ip}(\x)
=
\sum_{\ix}
\ec_{\ip}^{\ix}
u_{\ix}(\x)
\quad
\leadsto
\quad
\frac{\drm}{\drm\tT}
\ec_{\ip}^{\ix}
=
\sum_{\ipp}
\sum_{\ixp}
\big(\Q_{\ip\ipp}\big)_{\ix\ixp}
\ec_{\ipp}^{\ixp}
\;.
\end{equation}
Note that the apparent discreteness of this equation is not a
consequence of any approximation, e.g., grid discretisation.

Now, introducing appropriate vectors and matrices, the $2$-index
scalar recurrence can be cast into a vector recurrence in the
$\x$-basis index ($\Ntr$ is a large truncation index of the $\p$
basis)
\begin{equation}
\label{dcdt:matrix:x}
\mc_{\ix}
=
\left(
\begin{array}{c}
\ec_{0}^{\ix}
\\
\vdots
\\
\ec_{\Ntr}^{\ix}
\end{array}
\right)
\;,
\quad
\mQ_{\ix\ixp}
=
\left(
\begin{array}{ccc}
\big(\Q_{00}\big)_{\ix\ixp}
&
\cdots
&
\big(\Q_{0\Ntr}\big)_{\ix\ixp}
\\
\vdots
&
\ddots
&
\vdots
\\
\big(\Q_{\Ntr0}\big)_{\ix\ixp}
&
\cdots
&
\big(\Q_{\Ntr\Ntr}\big)_{\ix\ixp}
\end{array}
\right)
\;,
\quad
\frac{\drm}{\drm\tT}
\mc_{\ix}
=
\sum_{\ixp}
\mQ_{\ix\ixp}
\mc_{\ixp}
\;.
\end{equation}
To appreciate better the structure of the ``coefficients''
$\mQ_{\ix\ixp}$, their general expression is decomposed into the
``free'' and potential contributions $\mQ_{\ix\ixp}=\mQ_{\ix\ixp}^{\rm
f}+\mQ_{\ix\ixp}^{V}$.
For $\mQ_{\ix\ixp}^{\rm f}$ we find
\[
\mQ_{\ix\ixp}^{\rm f}
=
-
\left(
\begin{array}{ccccccc}
\gammao^{0}
\delta_{\ix\ixp}
\!&\!
\sqrt{1}
\Dp_{\ix\ixp}
\!&\!
\sqrt{1\!\cdot\!2}
\gammapp
\delta_{\ix\ixp}
\!&\!
0
\!&\!
0
\!&\!
0
\!&\!
\ddots
\\
\sqrt{1}
\Dm_{\ix\ixp}
\!&\!
\gammao^{1}
\delta_{\ix\ixp}
\!&\!
\sqrt{2}
\Dp_{\ix\ixp}
\!&\!
\sqrt{2\!\cdot\!3}
\gammapp
\delta_{\ix\ixp}
\!&\!
0
\!&\!
0
\!&\!
\ddots
\\
\sqrt{1\!\cdot\!2}
\gammamm
\delta_{\ix\ixp}
\!&\!
\sqrt{2}
\Dm_{\ix\ixp}
\!&\!
\gammao^{2}
\delta_{\ix\ixp}
\!&\!
\sqrt{3}
\Dp_{\ix\ixp}
\!&\!
\sqrt{3\!\cdot\!4}
\gammapp
\delta_{\ix\ixp}
\!&\!
0
\!&\!
\ddots
\\
0
\!&\!
\sqrt{2\!\cdot\!3}
\gammamm
\delta_{\ix\ixp}
\!&\!
\sqrt{3}
\Dm_{\ix\ixp}
\!&\!
\gammao^{3}
\delta_{\ix\ixp}
\!&\!
\sqrt{4}
\Dp_{\ix\ixp}
\!&\!
\sqrt{4\!\cdot\!5}
\gammapp
\delta_{\ix\ixp}
\!&\!
\ddots
\\
0
\!&\!
0
\!&\!
\sqrt{3\!\cdot\!4}
\gammamm
\delta_{\ix\ixp}
\!&\!
\sqrt{4}
\Dm_{\ix\ixp}
\!&\!
\gammao^{4}
\delta_{\ix\ixp}
\!&\!
\sqrt{5}
\Dp_{\ix\ixp}
\!&\!
\ddots
\\
0
\!&\!
0
\!&\!
0
\!&\!
\sqrt{4\!\cdot\!5}
\gammamm
\delta_{\ix\ixp}
\!&\!
\sqrt{5}
\Dm_{\ix\ixp}
\!&\!
\gammao^{5}
\delta_{\ix\ixp}
\!&\!
\ddots
\\
\ddots
\!&\!
\ddots
\!&\!
\ddots
\!&\!
\ddots
\!&\!
\ddots
\!&\!
\ddots
\!&\!
\ddots
\end{array}
\right)
\]
while the part due to $\V(\x)$ (including the quantum terms) has the
alternate dense structure:
\[
\mQ_{\ix\ixp}^{V}
=
-
\left(
\begin{array}{ccccccc}
0
\!\!\!&\!\!\!
[\G_{1}^{0,+}
\bV']_{\ix\ixp}
\!\!\!&\!\!\!
0
\!\!\!&\!\!\!
[\G_{3}^{1,+}
\bV^{(3)}]_{\ix\ixp}
\!\!\!&\!\!\!
0
\!\!\!&\!\!\!
[\G_{5}^{2,+}
\bV^{(5)}]_{\ix\ixp}
\!\!\!&\!\!\!
\ddots
\\[0.ex]
[\G_{1}^{0,-}
\bV']_{\ix\ixp}
\!\!\!&\!\!\!
0
\!\!\!&\!\!\!
[\G_{2}^{0,+}
\bV']_{\ix\ixp}
\!\!\!&\!\!\!
0
\!\!\!&\!\!\!
[\G_{4}^{1,+}
\bV^{(3)}]_{\ix\ixp}
\!\!\!&\!\!\!
0
\!\!\!&\!\!\!
\ddots
\\[0.ex]
0
\!\!\!&\!\!\!
[\G_{2}^{0,-}
\bV']_{\ix\ixp}
\!\!\!&\!\!\!
0
\!\!\!&\!\!\!
[\G_{3}^{0,+}
\bV']_{\ix\ixp}
\!\!\!&\!\!\!
0
\!\!\!&\!\!\!
[\G_{5}^{1,+}
\bV^{(3)}]_{\ix\ixp}
\!\!\!&\!\!\!
\ddots
\\[0.ex]
[\G_{3}^{1,-}
\bV^{(3)}]_{\ix\ixp}
\!\!\!&\!\!\!
0
\!\!\!&\!\!\!
[\G_{3}^{0,-}
\bV']_{\ix\ixp}
\!\!\!&\!\!\!
0
\!\!\!&\!\!\!
[\G_{4}^{0,+}
\bV']_{\ix\ixp}
\!\!\!&\!\!\!
0
\!\!\!&\!\!\!
\ddots
\\[0.ex]
0
\!\!\!&\!\!\!
[\G_{4}^{1,-}
\bV^{(3)}]_{\ix\ixp}
\!\!\!&\!\!\!
0
\!\!\!&\!\!\!
[\G_{4}^{0,-}
\bV']_{\ix\ixp}
\!\!\!&\!\!\!
0
\!\!\!&\!\!\!
[\G_{5}^{0,+}
\bV']_{\ix\ixp}
\!\!\!&\!\!\!
\ddots
\\[0.ex]
[\G_{5}^{2,-}
\bV^{(5)}]_{\ix\ixp}
\!\!\!&\!\!\!
0
\!\!\!&\!\!\!
[\G_{5}^{1,-}
\bV^{(3)}]_{\ix\ixp}
\!\!\!&\!\!\!
0
\!\!\!&\!\!\!
[\G_{5}^{0,-}
\bV']_{\ix\ixp}
\!\!\!&\!\!\!
0
\!\!\!&\!\!\!
\ddots
\\[0.ex]
\ddots
\!\!\!&\!\!\!
\ddots
\!\!\!&\!\!\!
\ddots
\!\!\!&\!\!\!
\ddots
\!\!\!&\!\!\!
\ddots
\!\!\!&\!\!\!
\ddots
\!\!\!&\!\!\!
\ddots
\end{array}
\right)
\]

The important point is that, for a given problem, with an appropriate
choice of the basis functions $\{u_{\ix}(\x)\}$, the matrix elements
in the indexes $(\ix,\ixp)$ may vanish when the second index lays at a
certain ``distance'' of the first.
Then, the recurrence relation (\ref{dcdt:matrix:x}) has a finite
coupling range, e.g.,
$\dot{\mc}_{\ix}
=
\mQ_{\ix,\ix-1}\mc_{\ix-1}
+
\mQ_{\ix,\ix}\mc_{\ix}
+
\mQ_{\ix,\ix+1}\mc_{\ix+1}$,
and can {\em always\/} be tackled by (matrix)
continued-fraction methods \cite{risken}.
Once the equations are solved, we can reconstruct the Wigner function
$\Wf(\x,\p)$ from the expansion coefficients $\ec_{\ip}^{\ix}$ and get
any observable from the general expression (\ref{average}) for the
averages.

\section{Application to quantum Brownian motion in periodic
  potentials}
In this problem one can anticipate that the coupling in $\ix$ will
coincide with the number of harmonics in $\V(\x)$: $1$ for cosine
potential ($3$-term recurrence), $2$ for a simple ratchet potential
($5$-term recurrence), etc.
Let us introduce the Fourier expansion of $\V'(\x)$ and plane waves as
basis functions:
%
\begin{equation}
\label{Vper}
\V'(\x)
=
\sum_{\ix}
\V_{\ix}'\,
\e^{\iu\ix\x}
\;,
\qquad
u_{\ix}(\x)
=
\frac{\e^{\iu\ix\x}}{\sqrt{2\pi}}
\;.
\end{equation}
It is convenient to extract the external force $\bFsh$ from $\V(\x)$
[including it in $\Dpm=\dplmi(\px-\bVo')-\etapm\bFsh$] and set the
auxiliary potential $\bVo$ proportional to the periodic part
$\bVo=\epspot\bV$.
Then, to obtain all matrix elements in $\mQ_{\ix\ixp}$ we only need
$(\px)_{\ix\ixp}=\iu\ix\,\delta_{\ix\ixp}$ and
$[\bV^{(\iq)}]_{\ix\ixp}=\V_{\ix-\ixp}^{(\iq)}$, whence
\begin{equation}
\label{DmDpVper:matr-elem}
\begin{array}{ccclccl}
\big[
\G_{\ip}^{\iq,\pm}
\bV^{(2\iq+1)}
\big]_{\ix\ixp}
&=&
&
&-&
\etapm^{2\iq+1}
G_{\ip}^{\iq}(-\etapro q^{2}\ldB^{2})
&
\V_{\ix-\ixp}'
\\
\Dpm_{\ix\ixp}
&=&
\big(\iu\,\dplmi\ix-\etapm\bFsh\big)
&
\delta_{\ix\ixp}
&-&
\dplmi\epspot
&
\bV_{\ix-\ixp}'
\end{array}
\end{equation}
Here $q=\ix-\ixp$ and
$G_{\ip}^{\iq}(z)
=
|\qcoef^{(\iq)}_{\ip}(q)|\exp(-z/2)\,\kum(-\ipp,2\iq+2\,;z)$
[cf.\ eq.~(\ref{Gamma:qcoeff})], being $\qcoef^{(\iq)}_{\ip}(q)$ the
previous quantum coefficient with $\ldB$ replaced by $q\ldB$.
Note that $\etapro=\etam\etap=\eta^{2}-1/4\leq0$.
Then the argument of $G_{\ip}^{\iq}$ is positive and $\exp(-z/2)$ acts
as a Gaussian regularisation factor [this is one advantage of the
inclusion of the parameter $\eta$ in eq.~(\ref{W:expansion})].
This factor reduces the weight of the off-diagonal terms inside
$\mQ_{\ix\ixp}$, contributing to the numerical stability and allowing
to study the deep quantum regime ($\ldB\gtrsim\xo$).

\begin{figure}
\threeimages[width=4.75cm]%
{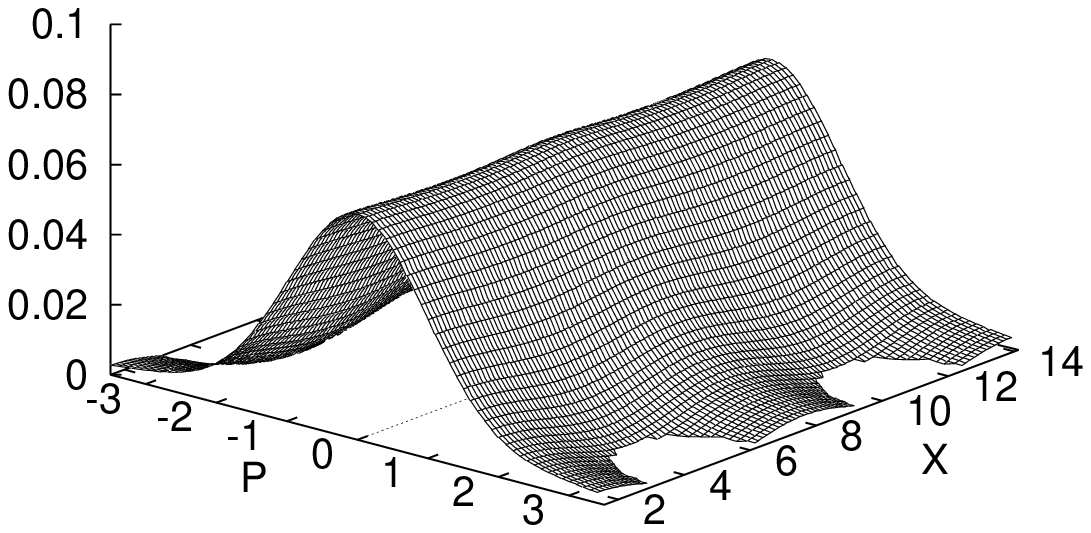}{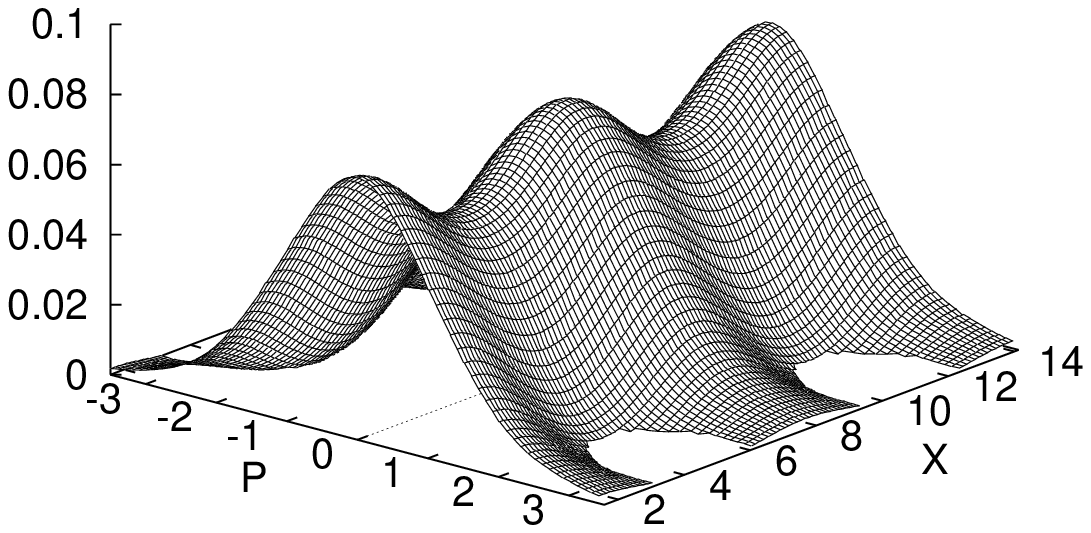}{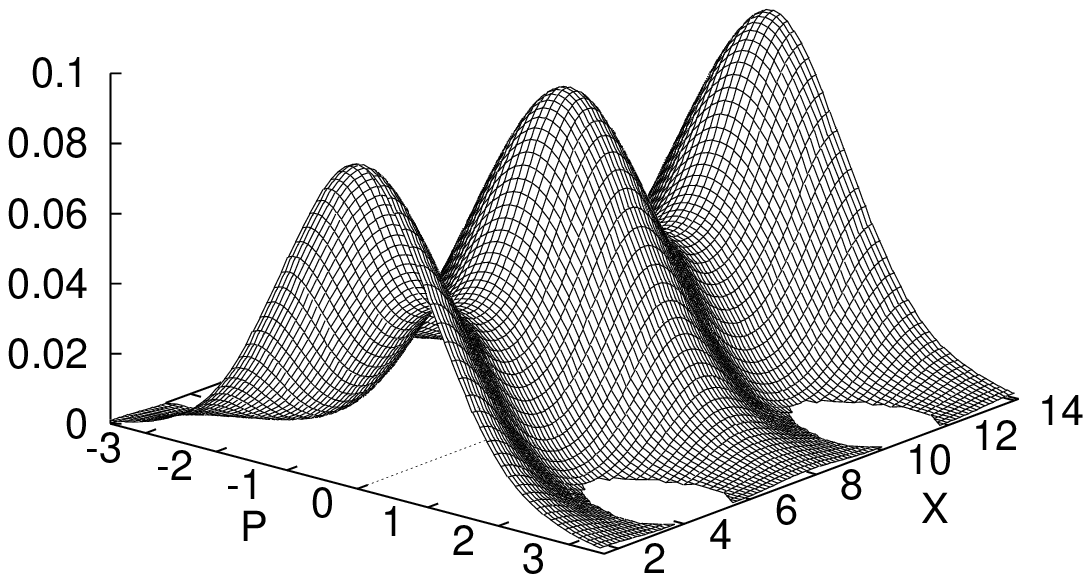}
\caption{
Wigner functions in the absence of dissipation for a particle in a
sinusoidal potential (one period and the two adjacent half-periods are
displayed).
The reduced Kondo parameter $\kondobar=2\pi(\So/\hbar)$ is $1$,
$1.57$, and $2.22$ (left to right), corresponding to $q=1/10$, $1/4$,
and $1/2$ of \cite[fig.~3]{kan98}; $q=(\kondobar/\pi)^{2}$.
}
\label{fig:kandemir}
\end{figure}

Let us give some examples.
Kandemir \cite{kan98} obtained the eigenfunctions of the Schr\"odinger
problem of a quantum particle in a sinusoidal potential in terms of
Mathieu functions.
They were represented with the Wigner function associated to the
density matrix of the pure states
$\rho(\x,\x')=\Psi(\x)\Psi^{\ast}(\x')$.
Setting the damping close to zero ($\gamma/\wo=10^{-6}$) we reobtain his
results for the ground state (fig.~\ref{fig:kandemir}).
It is nicely seen the evolution from delocalised solutions for the
lowest $\kondobar$ to localised ones as $\kondobar\propto\So/\hbar$
increases (the system becomes more ``classical'').

After checking the connection with the Hamiltonian limit, let us
include dissipation and temperature.
Chen and Lebowitz \cite{cheleb92} studied the transport
properties of underdamped particles in a cosine potential.
For low forces they obtained a free-particle like behaviour
$\langle\p\rangle\propto F/\gamma$.
Increasing $F$, the wave-vector associated to $\p$ reaches the first
zone boundary, where Bragg scattering reduces the velocity.
Eventually, for larger forces, $\p$'s corresponding to states in the
next band become available, and the free-particle behaviour is
progressively recovered.
\begin{figure}
\twoimages[width=7.25cm]{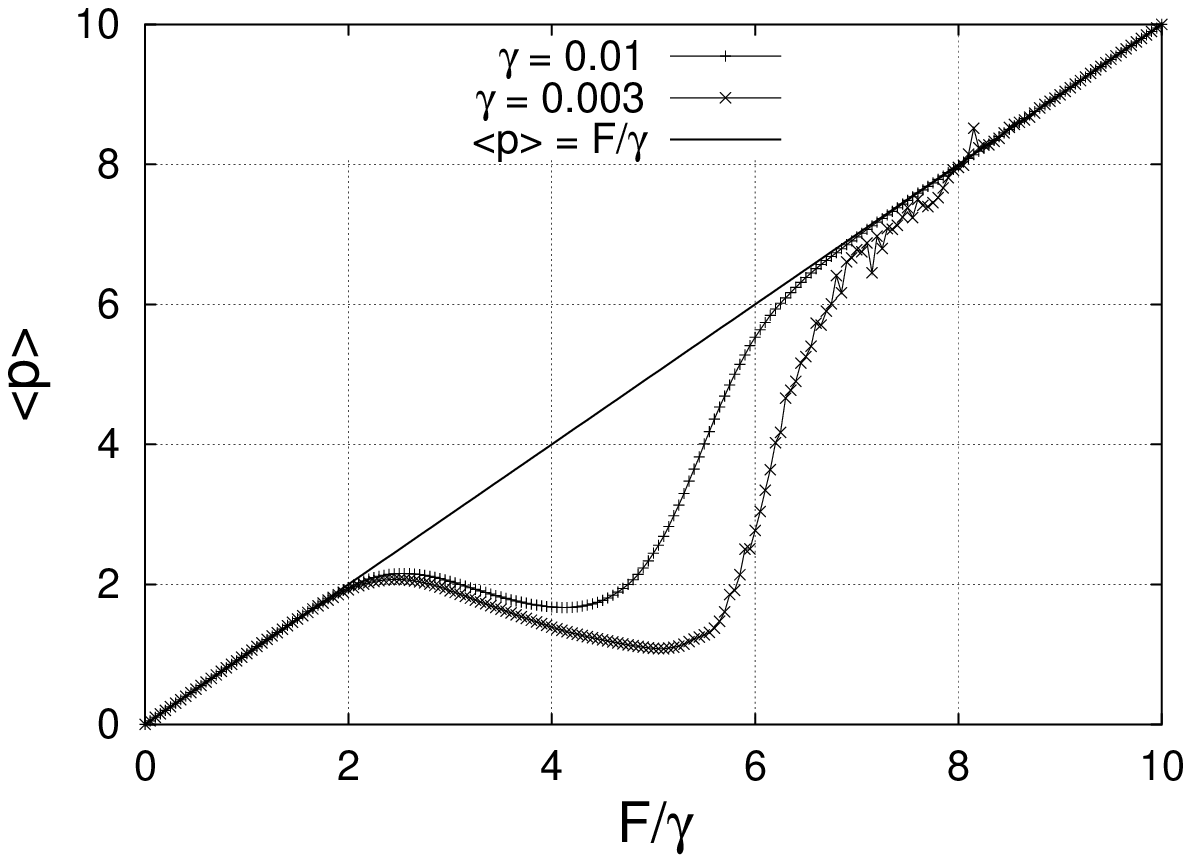}{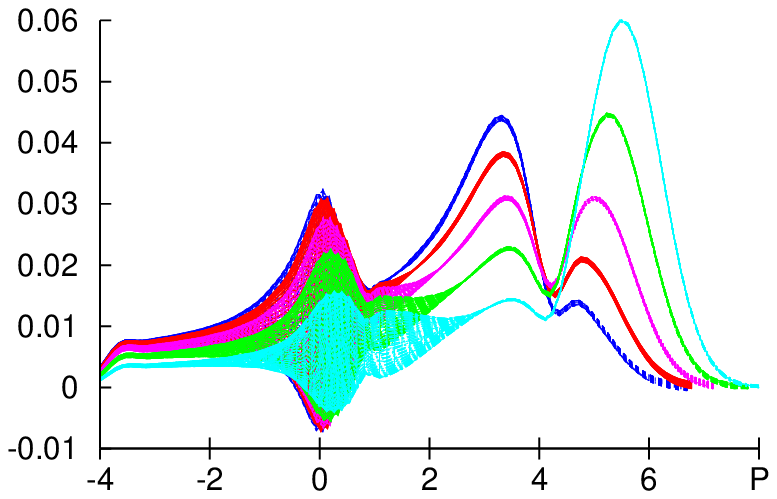}
\caption{
Left panel: $\langle\p\rangle$ vs.\ force  for underdamped
particles at $T=0.5$.
The reduced Kondo parameter is $\kondobar=0.7854$, which corresponds
to $\Omega_{q}=4$ of \cite{cheleb92}; $\Omega_{q}=\pi/\kondobar$.
Right panel: Side view of the Wigner function ($\Wf$ vs.\ $\p$,
$\forall x$) for $\gamma=0.01$ at $F/\gamma=4.5$, $4.75$, $5$, $5.25$,
and $5.5$ (left to right).
}
\label{fig:cheleb92}
\end{figure}

Since high $T/\gamma$ approximations are involved in
\cite{cheleb92}, we set $\Dpp=\gamma\sM\kT$ and $\Dxp=0$ in
the master equation (\ref{WKK:Dxp}).
Solving it with the continued-fraction method we reobtain the effect
described above (see fig.~\ref{fig:cheleb92}).
However, with this method we also get the Wigner distribution.
Representing the obtained $\Wf$ vs.\ $\p$ in the range where the
curves start to rise again ($F/\gamma\sim5$), we find two peaks at
values associated roughly to the lowest bands ($\p\sim2$ and $6$),
supporting their interpretation.
In analogy with  the classical case, we would  speak of multistability
between the  ``locked'' ($\p\sim0$) and {\em  two} quantum ``running''
solutions.
In the classical case the running solutions have a wiggling structure
along $\x$ \cite[fig.~11.22]{risken} (the particle slows down near the
potential maxima).
Displaying the side view of $\Wf$, instead of the true marginal
distribution $P(\p)=\int\!\drm{\x}\,\Wf(\x,\p)$, we can appreciate the
straight structure (substrate insensitive) of the {\em quantum\/}
running solutions.

\begin{figure}
\twoimages[width=7.25cm]{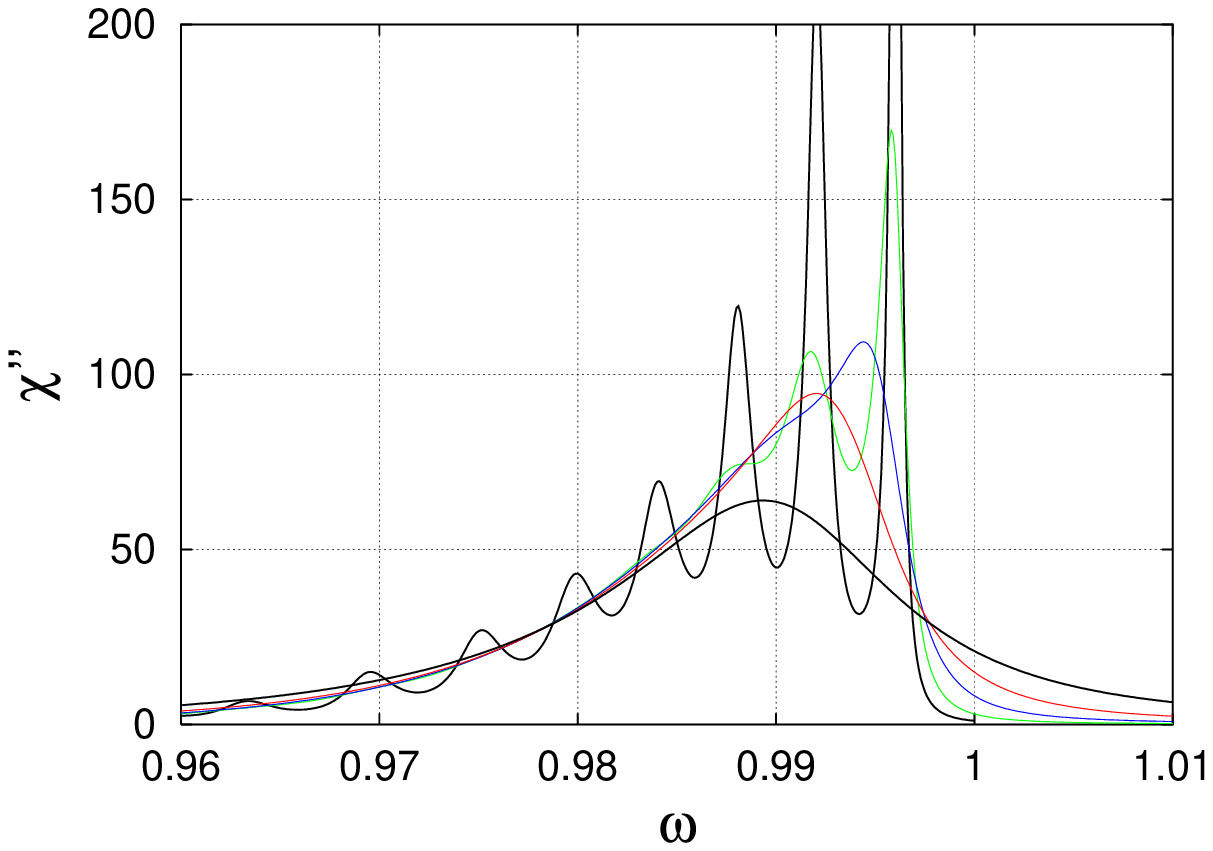}{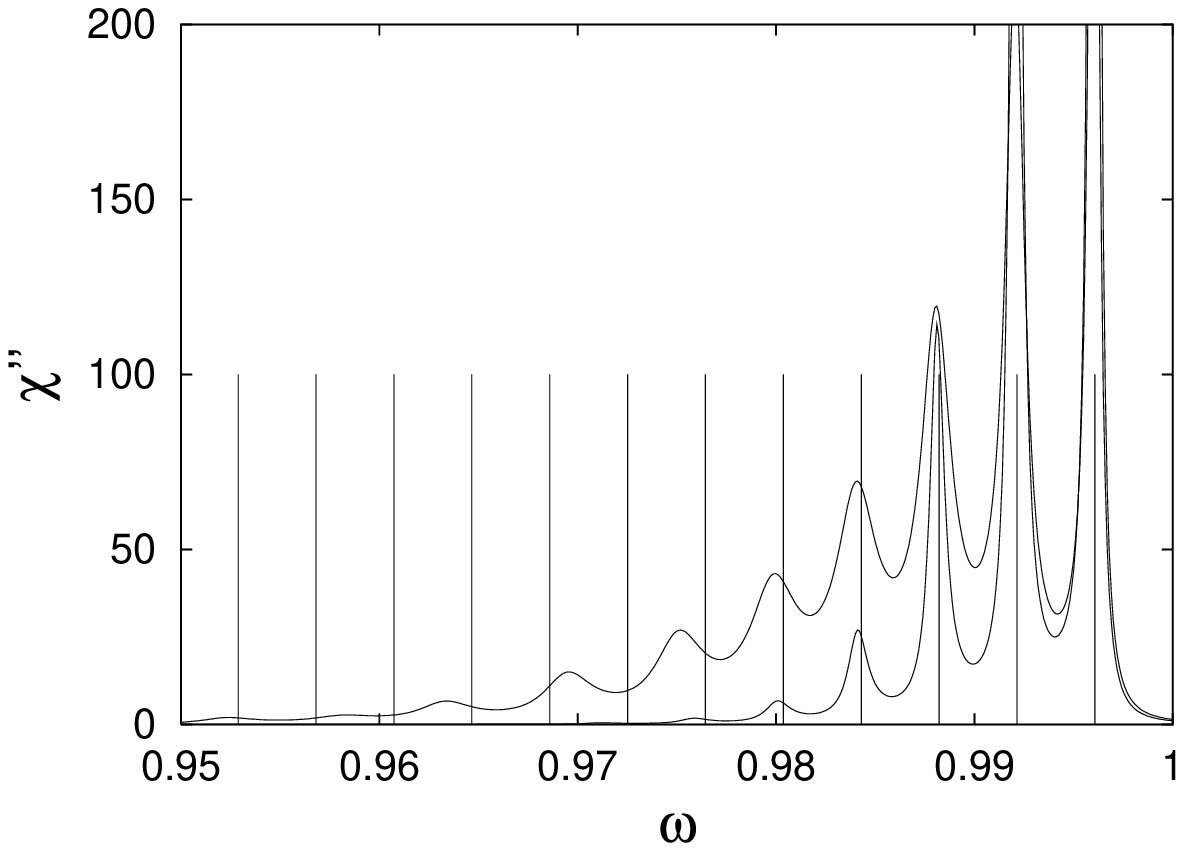}
\caption{
Imaginary part of the dynamical susceptibility.
Left panel: $T=0.05$ and $\kondobar=200$ with $\gamma=0.01$, ,
$0.003$, $0.001$, $0.0003$, $0.0001$ (bottom to top).
Right panel: $\gamma=0.0001$ with $T=0.05$ and $T=0.025$ (lower curve).
Vertical lines: transition frequencies of the associated anharmonic
oscillator.
}
\label{fig:suscep-w}
\end{figure}

Before concluding let us consider an example of dynamical response.
Figure~\ref{fig:suscep-w} shows the linear dynamical susceptibility
vs.\ frequency of the ac field of a particle in a sinusoidal
potential.
In the classical case, the line-shape $\chi''(\omega)$ broadens and
extends to $\omega$'s lower than the frequency of oscillation near the
bottom of the potential wells ($\omega=1$ in scaled units).
The reason is the dependence of the oscillation period on the
amplitude of oscillation in anharmonic potentials, which yields a
non-dissipative contribution to $\chi''(\omega)$ \cite{risken}.

In the quantum regime (approached by decreasing the Kondo parameter
$\kondo\propto\gamma\kondobar$) the $\chi''(\omega)$ curves develop a
multipeaked structure, which could be loosely interpreted as the
nonlinear oscillations becoming quantised.
We have calculated analytically the effect of the anharmonic terms of
a cosine potential (those beyond $\propto\x^{2}$) on the energy levels
of the harmonic oscillator part.
Doing this with first-order perturbation theory, we find the
dependence on the energy-level index $k$ acquired by the distance
between levels $\Delta E_{k+1,k}$.
The corresponding frequencies $\Delta E_{k+1,k}/\hbar$ agree quite
well with the location of the main peaks (fig.~\ref{fig:suscep-w},
right panel), substantiating the above interpretation.
Finally, when the temperature is lowered the thermal population of the
higher levels is reduced.
Then, the transitions between these levels (``large amplitude
oscillations'') become less probable and the corresponding peaks
reduced.

\section{Discussion}
We have adapted the continued-fraction method of classical Brownian
motion to tackle quantum master equations in phase space.
As the classical counterpart, the method is limited to a few degrees
of freedom.
Besides, it is more problem dependent, as it uses the $\x$-recurrence
and requires a good position basis.
Nevertheless, it inherits most advantages from the classical method:
efficiency, accuracy, and the obtaining of the ``distribution''
$\Wf(\x,\p)$, from which {\em any\/} observable can be computed.
Besides, it is essentially nonperturbative and specially apt for
stationary solutions (static and dynamic).
The eigenvalue spectrum is not required, which is desirable when
dealing with non-bounded Hamiltonians, continuous spectra, etc.
The connection with the classical results is always possible and
attained in a natural way.
Indeed, the knowledge of the classical phase-space structure
(separatrices, attractors, etc.) plus the visualization of the
Wigner function, can provide valuable insight into complex problems
involving nonlinear quantum systems subjected to dissipation and
fluctuations.

\acknowledgments
This work was supported by DGES (Spain), project BFM2002-00113.
Valuable discussions with F.\ Falo and D.\ Zueco are gratefully
acknowledged.

\end{document}